\documentclass{article}
\usepackage{dcase2024_techrep,amsmath,graphicx,url,times,booktabs, tabularx}
\usepackage{textgreek}
\usepackage{float}
\usepackage{amsmath}
\usepackage{amssymb}
\usepackage{graphicx}

\title{ResNet-Conformer Network with Shared Weights and Attention Mechanism for Sound Event Localization, Detection, and Distance Estimation}

\name{Quoc Thinh Vo, David K. Han\thanks{Thanks to the Office of Naval Research for funding.}}
\address{Author Affiliation(s)}

\address{Drexel University, College of Engineering\\
      Electrical and Computer Engineering Department\\
      3100 Market St, Philadelphia, PA 19104, USA\\
      {qv23, dkh42}@drexel.edu}

\begin{document}

\ninept
\maketitle

\begin{sloppy}

\begin{abstract}
This technical report outlines our approach to Task 3A of the Detection and Classification of Acoustic Scenes and Events (DCASE) 2024, focusing on Sound Event Localization and Detection (SELD). SELD provides valuable insights by estimating sound event localization and detection, aiding in various machine cognition tasks such as environmental inference, navigation, and other sound localization-related applications.

This year's challenge evaluates models using either audio-only (Track A) or audiovisual (Track B) inputs on annotated recordings of real sound scenes. A notable change this year is the introduction of distance estimation, with evaluation metrics adjusted accordingly for a comprehensive assessment.

Our submission is for Task A of the Challenge, which focuses on the audio-only track. Our approach utilizes log-mel spectrograms, intensity vectors, and employs multiple data augmentations. We proposed an EINV2-based \cite{cao2021improved} network architecture, achieving improved results: an F-score of 40.2\%, Angular Error (DOA) of 17.7$^\circ$, and Relative Distance Error (RDE) of 0.32 on the test set of the Development Dataset \cite{Politis2022starss22, Shimada2023starss23}.

\end{abstract}

\begin{keywords}
 log-mel spectrogram, sound event detection and localization, distance estimation, attention mechanism
\end{keywords}

\section{Introduction}
\label{sec:intro}

The goal of the Sound Event Localization and Detection (SELD) task is to detect sound events (SED) while estimating their corresponding direction of arrival (DOA). SELD systems have shown significant potential in diverse applications such as machine listening, smart homes, navigation, and wildlife sound detection. The annual DCASE challenge has drawn extensive attention from researchers and has facilitated notable advancements in the field of SELD.

In the current field, research aimed at solving the SELD problem can be mainly classified into two main approaches. The first category includes a model architecture featuring a unified input and output system. In the DCASE 2020 challenge, the activity-coupled Cartesian DOA (ACCDOA) representation was introduced in \cite{shimada2021accdoa}. This representation maps sound event activity to the length of a corresponding Cartesian DOA vector, effectively merging the SED and DOA tasks into a single regression task within Cartesian coordinates. However, the ACCDOA representation has limitations in dealing with simultaneous similar events. To overcome this, the ACCDOA representation was enhanced to multi-ACCDOA by incorporating Auxiliary Duplicating Permutation Invariant Training (ADPIT) \cite{liu2018permutation} as discussed in localizing and
detecting overlapping sounds from the same class \cite{shimada2022multi}. The multi-ACCDOA output has been used as the standard in the Baseline systems for the DCASE 2023, and 2024 challenges.

The second approach is a two-branch structure model. In 2019, a two-step strategy was proposed in the proposed methods like \cite{mazzon2019sound}, Cao et al.'s system \cite{cao2019polyphonic} uses a logmel magnitude spectrogram with M = 96 mel bins and the generalized cross-correlation phase transform (GCC-PHAT) \cite{knapp1976generalized}. Since the logmel spectrum lacks phase information important for DOA estimation, the author \cite{cao2019polyphonic} used GCC-PHAT as additional acoustic features. This method employs a two-stage training approach: initially training only the SED branch of the network, followed by transferring the parameters of the Convolutional Neural Network (CNN) blocks responsible for computing high-level features to the DOA branch for separate training. During DOA branch training, SED ground truth labels mask the estimated DOA labels. During inference, both SED and DOA are predicted using the independently trained branches, with DOA labels adjusted by the predicted SED labels. This strategy streamlines training while utilizing SED features for DOA estimation. Additionally, the CNN architecture diverges from the Baseline system, notably employing a 2x2 pooling layer that compresses features along the time axis, followed by up-sampling at the conclusion. Furthermore, an event-independent network version 2 (EINV2) was introduced in \cite{cao2021improved}, which incorporates soft parameter sharing and multi-head self-attention (MHSA) to decode the SELD outputs effectively.

Both approaches demonstrate notable performance improvements in the SELD task. Hence, we aim to leverage the EINV2-based design with multi-ACCDOA output to capitalize on the strengths of both systems.

To fully leverage the time-frequency features of audio data, we devised a method integrating diverse techniques for data augmentation and feature extraction. Our approach involved designing and training a novel network within the EINV2 framework. This included applying various time-frequency domain augmentation techniques to enrich training data diversity and enhance model robustness. Additionally, we incorporated a multi-scale channel attention mechanism to effectively capture inter-channel correlation information and employed multi-phase training to optimize model performance through domain-specific training strategies.

In summary, the main contributions of our proposed method are:

We designed a network architecture based on the EINV2 framework specifically designed for the SELD task.

Our approach includes a split-phase training framework and employs diverse data augmentation techniques such as random cutout \cite{zhong2020random}, noise injection, SpecAugment \cite{park2019specaugment}, and Audio Channel Swapping (ACS) \cite{wang2023four} to enhance model generalization and performance.

We incorporated a multi-scale channel attention mechanism to capture inter-channel correlations effectively, enhancing the model's ability to handle overlapping sound events across different classes, leveraging Conformer \cite{peng2021conformer} integration.

Experimental results on the Development Dataset demonstrate significant improvements compared to the Baseline system, demonstrating the effectiveness of our approach in tackling the challenges of SELD.

\section{PROPOSED METHODOLOGY}
\label{sec:methodology}

\begin{figure*}[htb]
\centering
\includegraphics[width=\textwidth]{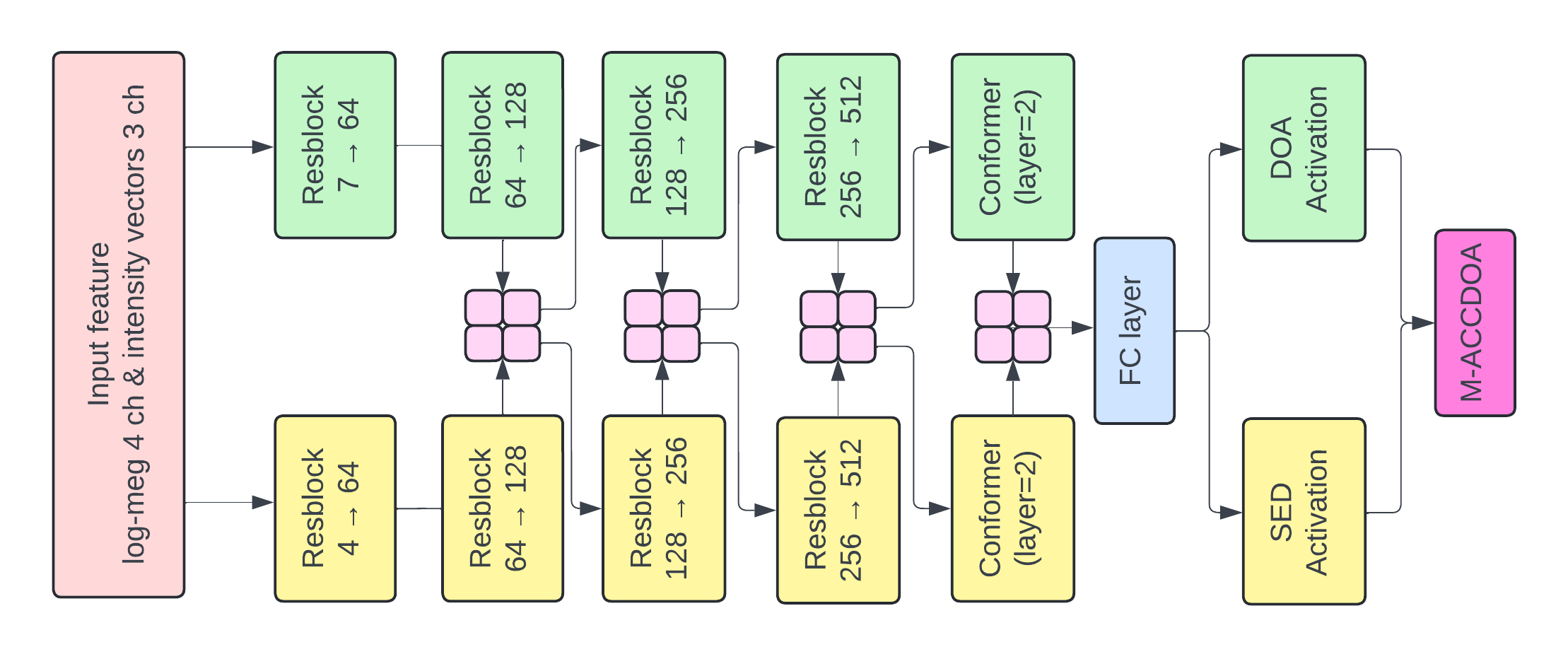}
\caption{Proposed Network on effective training of the SELD task}
\label{fig:proposed_method}
\end{figure*}

\subsection{Features}
In our approach, we use audio files in Ambisonic format of the Development Dataset 2024 \cite{Shimada2023starss23}. The dataset comprises 7 hours and 22 minutes of real recordings, divided into 90 training clips and 78 testing clips. It includes 13 sound event classes: female speech, male speech, clapping, telephone, laughter, domestic sounds, footsteps, door, music, musical instrument, water tap, bell, and knock. The scenarios involve up to three overlapping sound sources. We extracted two types of features from the audio files: 4-channel spectrograms were accumulated into 64 mel energies, and 3-channel sound intensity vectors, as detailed in \cite{yasuda2020sound}. These features were used as inputs for our proposed network. The input feature matrix, with dimensions C × T × F, where C denotes channels, T represents frame sequence length, and F indicates feature count, was fed into the model. The audio data was sampling at 24 kHz.

\subsection{Data Augmentation} 
To improve the model's performance and generalization, we apply data augmentation in both time and frequency domains. Techniques include random cutout and noise injection. Moreover, spectrogram augmentation is implemented using SpecAugment \cite{park2019specaugment}, a widely adopted method from related experiments in the domain.

In this year's challenge, the provided dataset, Sony-TAu Realistic Spatial Soundscapes 2023 (STARSS23), has expanded since its inception. However, enhancing model robustness remains a challenge. To address this, we augmented the dataset obtained by generating synthetic data using SpatialScaper \cite{Roman2024} and applied the ACS technique \cite{wang2023four} in the final stage of training, thereby increasing the dataset size eight-fold. Additionally, we continue to employ techniques such as random cutout, time-frequency masking, and frequency shifting to improve model generalization in both stages of the training. Finally, we introduced random mix to blend original and augmented data with specified weights, creating a new training dataset.

\subsection{Architecture} 
We propose a ResNet-Conformer model based on the EINV-2 \cite{cao2021improved} architecture and the EINV2-based with MS-CAM \cite{dai2021attentional} blocks from Xue et al. \cite{xueattention}. Our approach enhances the Baseline model with several modifications: integrating ResNet blocks with rescaled residual connections for improved performance, incorporating a multi-scale channel attention mechanism to fuse local and global features across channels, and replacing GRU and MHSA with Conformers \cite{peng2021conformer} to better capture temporal features. Furthermore, we introduce max pooling to mitigate overfitting after each shared weighted layer. Leveraging the EINV2 framework's success in SELD tasks, our ResNet-Conformer integrates ResNet blocks with the EINV2 two-branch design featuring shared weights. Unlike the EINV2's multi-branch output, our model adopts a multi-ACCDOA format. See Figure \ref{fig:proposed_method} for an illustration of our network structure.

\subsection{Training}

We trained the model using back-propagation and the Adam optimizer with a batch size of 512. The output is in multi-ACCDOA format with the MSE-ADPIT loss function as described in \cite{krause2024sound}.

The modification of the single-task multi-ACCDOA approach introduced in \cite{shimada2022multi} expands the original 3-element DOA vector to include an additional distance estimate. For \(N\) tracks, \(C\) classes, and \(T\) frames, the output is defined as \(y_{nct} = [a_{nct}R_{nct}, D_{nct}]\), where \(n\), \(c\), and \(t\) represent the output track number, target class, and time frame, respectively. In this context, \(a_{nct} \in \{0, 1\}\) indicates detection activity, \(R_{nct} \in \langle -1, 1 \rangle\) refers to the DOA vectors, and \(D_{nct} \in \langle 0, \infty)\) denotes distance values. The dimensions are specified as follows: \(a, D \in \mathbb{R}^{N \times C \times T}\), \(R \in \mathbb{R}^{3 \times N \times C \times T}\), and \(\|R_{nct}\| = 1\). As modeled by Krause et al. \cite{krause2024sound}, up to \(N = 3\) is considered, resulting in number of output neurons \(= 156\). The output is linear to cover both DOA and distance ranges. The final loss function is defined as follows \cite{krause2024sound}:

\begin{equation}
L^{\text{ADPIT}} = \frac{1}{CT} \sum_{c}^{C} \sum_{t}^{T} \min_{\alpha \in \text{Perm}[ct]} l_{\alpha,ct}^{\text{ACCDOA}}
\end{equation}

\begin{equation}
l^{\text{ACCDOA}}_{\alpha,ct} = \frac{1}{N} \sum_{n}^{N} L(y_{\alpha,nct}, \hat{y}_{\alpha,nct})
\end{equation}

where \(L(\cdot)\) is a chosen loss function, \(\alpha\) is one possible track permutation, and \(\text{Perm}[ct]\) is the set of all possible permutations.

A learning rate scheduler and early stopping were used during training to prevent overfitting. The model was trained in parallel mode using three NVIDIA GeForce RTX 3090 GPUs. Additionally, one of our submissions employed a full training duration of 120 epochs.

We also adopt a multi-phase training strategy. Initially, the model's weights are initialized with synthetic data \cite{krause2024dcase}. Subsequently, we fine-tune the model using the Development Dataset 2024, incorporating ACS and the augmented data to generate more data. This combined augmentation and training strategy substantially enhances both the robustness and performance of the model.

\subsection{Evaluation metrics}
To evaluate our models, we apply SELD metrics defined in the DCASE Challenge 2024 Task 3, including F-score for SED, Angular Error, and Relative Distance Error. Detection metrics consider spatial proximity, with F-score (F20°) requiring correct predictions only if the class matches, Angular Error is \(\leq 20^\circ\), and Relative Distance Error is \(\leq 1.0\). The Relative Distance Error measures the difference between estimated and reference distances normalized by the reference distance itself. Evaluation is conducted in one-second segments using micro-averaging, and source matching employs the Hungarian algorithm based on angular distance \cite{politis2020overview}.

\section{Experimental Results}
\label{sec:experimentalresults}

Table 1 presents our results on the Development Dataset 2024. The network was trained using a multi-phase approach involving synthetic, real, and augmented data.

Model 1, trained without the multi-phase training framework or data augmentation, achieved an F-score of 23.1\%, DOA of 25.3$^\circ$, and RDE of 0.33.

For Model 2, we initialized weights with synthetic data and subsequently trained on the real dataset with augmented data, excluding ACS. Model 2 achieved an F-score of 33.0\%, DOA of 20.7$^\circ$, and RDE of 0.32.

The Proposed Model utilized weight initialization from synthetic data and was trained on real data with ACS and random mix augmentations to significantly enhance and diversify training data. It achieved an F-score of 40.2\%, DOA of 17.5$^\circ$, and RDE of 0.32. This model was used for inference and submitted for the DCASE 2024 Task 3A challenge.

\begin{table}[htp]
\begin{tabular}{|l|l|l|l|}
\hline
\textbf{Model } & \textbf{ F$_{20^\circ}$} & \textbf{ DOA$_{\text{CD}}$} & \textbf{ RDE$_{\text{CD}}$}\cr
\hline
Baseline & 13.1\% & 36.90$^\circ$ & 0.33 \cr
Model 1 & 23.1\% & 25.3$^\circ$ & 0.33 \cr
Model 2 & 33.0\% & 20.7$^\circ$ & 0.32 \cr
\textbf{Proposed Model} & \textbf{40.2\%} & \textbf{17.5$^\circ$} & \textbf{0.32} \cr

\hline
\end{tabular}
\caption{Reported metrics for the test on the Development Dataset 2024}
\label{tab:table1}
\end{table}

\section{CONCLUSION and FUTURE WORK}
\label{sec:conclusion}
In this experiment, we implemented a ResNet-Conformer two-branch network with multi-phase training, achieving improved performance compared to the Baseline system in terms of F-score and Angular Error.

However, our results indicate that the Relative Distance Error, a new evaluation metric introduced for this year's challenge task, did not show significant improvement over the Baseline. Future research efforts will focus on enhancing performance on this metric.

\section{ACKNOWLEDGMENT}
\label{sec:ack}

The work on this paper is sponsored by the Office of Naval
Research (Grant Number: N00014-21-1-2790)

\bibliographystyle{IEEEtran}
\bibliography{refs}
%
%
%
%
%
%
%
%
%

\end{sloppy}
\end{document}